\documentclass[conference]{IEEEtran}
\usepackage{amsmath}

\usepackage{subcaption}
\usepackage{framed}
\usepackage{color,soul}
\usepackage{nameref}
\usepackage{balance}       
\usepackage{graphics}      
\usepackage[T1]{fontenc}   
\usepackage{txfonts}
\usepackage{mathptmx}
\usepackage[pdflang={en-US},pdftex]{hyperref}
\usepackage{color}
\usepackage{booktabs}
\usepackage{textcomp}

\usepackage{microtype}        
\usepackage{ccicons}          

\usepackage{todonotes}

\title{Developing and Deploying a Taxi Price Comparison Mobile App in the Wild: Insights and Challenges}

\author{\IEEEauthorblockN{Anastasios Noulas\IEEEauthorrefmark{1},
Vsevolod Salnikov\IEEEauthorrefmark{2},
Desislava Hristova\IEEEauthorrefmark{3}, 
Cecilia Mascolo\IEEEauthorrefmark{3} and
Renaud Lambiotte\IEEEauthorrefmark{2}}
\IEEEauthorblockA{\IEEEauthorrefmark{1}Data Science Institute\\
Lancaster University\\ 
a.noulas@lancaster.ac.uk}
\IEEEauthorblockA{\IEEEauthorrefmark{2}Department of Mathematics\\
University of Namur\\
name.surname@unamur.be}
\IEEEauthorblockA{\IEEEauthorrefmark{3}Computer Laboratory\\
University of Cambridge\\
name.surname@cl.cam.ac.uk}
}

\begin{document}

\maketitle

\begin{abstract}
As modern transportation systems become more complex, there is need for mobile applications that allow travelers to navigate efficiently in cities. In taxi transport the recent proliferation of Uber has introduced  new norms including a flexible pricing scheme where journey costs can change rapidly depending on passenger demand and driver supply. To make informed choices on the most appropriate provider for their journeys, travelers need access to knowledge about provider pricing in real time. To this end, we developed OpenStreetcab a mobile application that offers advice on taxi transport comparing provider prices. We describe its development and deployment in two cities, London and New York, and analyse thousands of user journey queries to compare the price patterns of Uber against major local taxi providers. We have observed large heterogeneity across the taxi transport markets in the two cities. This motivated us to perform a  
price validation and measurement experiment on the ground comparing Uber and Black Cabs in London. The experimental results reveal interesting insights: not only they confirm feedback on pricing and service quality received by professional drivers users, but also they reveal the tradeoffs between prices and journey times between taxi providers. With respect to journey times in particular, we show how  experienced taxi drivers, in the majority of the cases, are able to navigate faster to a destination compared to drivers who rely on modern navigation systems. We provide evidence that this advantage becomes stronger in the centre of a city where urban density is high. 

\end{abstract}

\section{Introduction}
\label{sec:introduction}
The development of ubiquitous location sensing technologies and the resulting availability of data layers of human mobility in urban transport and road networks have enabled the proliferation of urban transport mobile apps. These systems are further fueled by the increase in APIs provided by transport authorities~\cite{tflapi,mtaapi} or, in certain cases, hacktivists publishing online large amounts of mobility datasets that were previously inaccessible, stored away in old devices of public organisations~\cite{Foiling,RealTimeTrains}.
The focus of this wave of apps has been primarily on assisting citizens with navigation in rail or bus transportation systems. It is, to a large extent, the growing complexity of these urban systems~\cite{gallotti2016lost} that has brought forward the necessity for such intelligent solutions. Some of these are now exploited by millions of users globally so as to navigate urban environments efficiently by minimizing financial costs and journey time duration~\cite{MapWay,CityMapper}. However, there is little knowledge on tools to cater for the increasing complexity of taxi provider selection. 

The necessity for making intelligent choices as we travel
has risen not only from the fact that typically a large number of providers operate in the same geographic space, but also due to the large temporal variability in the quality of services offered, as well as in prices.
With respect to taxi transport specifically, tariff-based 
prices have been traditionally in place, which imply standard costs per mile and per second travelled.
Despite the existence of tariffs, however, compared to fixed-line transportation systems, those that are based on vehicle movement are inherently harder to track due to variations in travel times driven by urban congestion or alternative routes picked by drivers~\cite{skog2009car}. Hence, the exact price a customer would pay  is not easily predictable ahead of a journey's start time.  More recently, the new pricing scheme introduced in the industry by Uber, popularly known as surge pricing~\cite{surgepricingtime,surgepricingslate}, has made the choice of  the cheapest taxi provider even more complex. 
Prices change in real time in accordance to passenger demand and driver supply.
What is more, in comparison to aviation and flight search services online, in the case of taxi transport, users will typically need to access information on pricing on the move and in real time.

In response to the growing complexity of taxi transport dynamics, which affects a growing number of cities around the world~\cite{ubercities}, 
we describe the process of development of OpenStreetCab, a mobile application that aims to assist users in choosing a taxi provider in a city in real time, offering estimates on taxi prices. 
We reflect on our design decisions and discuss the application's
usage and pricing statistics between two cities and two taxi providers. 
We provide a validation of the app's price estimates and a comprehensive study of price and journey time measurement through a real world experiment which compared taxi providers in the city of London.  

More specifically we make the following contributions: 
\begin{itemize}
\item We describe the development and refinement process of OpenStreetCab available on Android and iOS that provides taxi journey price estimates to users in real time, given as input their journey's origin and destination. 
\item Through the app, we collect a dataset on origin/destination price queries, generated by thousands of users that have used it in London and New York. By conducting an analysis of user queries in the two cities, we observe variations in terms of how Uber's cheapest service, Uber X, compares to the local cab companies in each city.
For example, Uber X tends to be more expensive on average than Yellow Cabs in New York, but the same is not true for Black Cabs in London.  
\item Motivated by the data driven and user insights we acquire, we performed a set of experiments on the ground in order to validate the price estimates provided by OpenStreetCab, but also to understand routing behavior and measure journey timings of the two taxi services in London. 
\item We show how integrating feedback in the application's logic leads to better price estimates
and alleviates systematic inaccuracies on the prediction of routes and their corresponding driving times provided by  pricing APIs.
\item We demonstrate how professional and trained taxi drivers present a better routing skill in a dense and complex urban environment where computer navigation systems struggle. 
Drivers are more likely to pick side streets which can potentially help them navigate away from traffic, especially within the urban core of the city where street, place and vehicle density maximise. 
This advantage however is being progressively lost as we move to the more sparsely populated and larger in area size urban outskirts.
\end{itemize}
These results highlight not only the trade-offs between pricing and journey durations in taxi mobility, that could be taken into account by related applications and services, but additionally, they reveal interesting differences between human and computer-assisted routing in urban environments. 
Overall, our findings are relevant for mobile developers and researchers active in the domain
of urban transport. 


\section{Related Work}
\label{sec:related}
Digital traces of travelers in transportation systems such as underground rail and bus networks have been heavily employed to study travel behavior and suggest better travel routines and improvements~\cite{agard2006mining,lathia2011mining,o2014mining,foell2013mining}. 

At the same time, mobile applications which aim at easing travelling experience have become increasingly popular. Google Maps has, aside from routing support via car or public transport, added an Uber integration feature, through which Uber users can search for a destination on Google Maps and directly receive information on journey times as well as costs~\cite{googlemapsintegration}. CityMapper~\cite{CityMapper} and MapWay~\cite{MapWay} are offering even more specialized information on transport options with respect to the requested route.
For instance, CityMapper now reports the expected number of calories a person would burn when navigating through a particular route, given a transport mode (e.g. bike vs walk). Price comparison services exists for flights or trains.
However these services, even if provided with apps, work at a much slower timescale than any city taxi or urban transport services: the booking of long distance train and flight is usually done days if not months in advance,  allowing ample time for the system to learn trends and apply corrections.

Cycle sharing networks have been also extensively studied,
popularized in many cities as a sustainable mode of transport~\cite{kaltenbrunner2010urban}. 
In these systems, the pricing model is usually flat, with standard charges on a per hour basis. Cycles can only be picked up from specific locations, making the price estimation problem easier. None of the systems analyzed have more than one providers for the service, which also simplifies the problem.

In terms of taxi studies, some work have mined the mobility
trajectories of taxis~\cite{yuan2010t,zheng2011computing} with applications in route discovery, activity recognition or privacy aware mobility models to name a few examples. In ubiquitous computing, applications have been powered by the analysis of datasets that describe taxi trajectories. For example Zheng et. al in~\cite{zheng2011computing}
analyze taxi data to identify regions with traffic problems and correlations amongst geographic areas in terms of taxi mobility to assess the effectiveness of urban planning projects. The modeling of taxi sharing, otherwise known as taxi pooling, schemes has been another subject of study~\cite{santi2014quantifying} due to its potential in relieving cities from traffic congestion. Routing behavior of drivers and its relationship  with navigation systems and traffic congestion has been a related topic of study~\cite{lima2016understanding} on vehicle and taxi movement. 
Our work is partly related to taxi trajectories analysis~\cite{zheng2011urban}, however mainly in relationship to the goal of understanding how we can improve the information we give to users. 
Finally, in terms of taxi mobile applications, there are numerous that have appeared in mobile marketplaces offering taxi booking services~\cite{lyft,uber,hailo} some of which provide information
on the costs and other characteristics of taxi providers~\cite{Way2Ride}. In~\cite{noulas2015mining} we discussed aspects of the spatio-temporal dynamics of surge pricing in the context of OpenStreetCab and used mobility data external to Uber to predict surge across geographic areas in New York. Our goal in this work is to provide insights on taxi price comparison focusing on the deployment in two cities (Sections~\ref{sec:application}~and~\ref{sec:analysis}). Further, by comparing providers through a measurement driven experiment on the ground, we identify critical aspects in routing behavior that can help better estimating time and prices (Section ~\ref{sec:experiment}). 

\begin{figure}[t!]
    \centering
    \begin{subfigure}[b]{.5\textwidth}
  \centering
    \includegraphics[width=0.8\columnwidth]{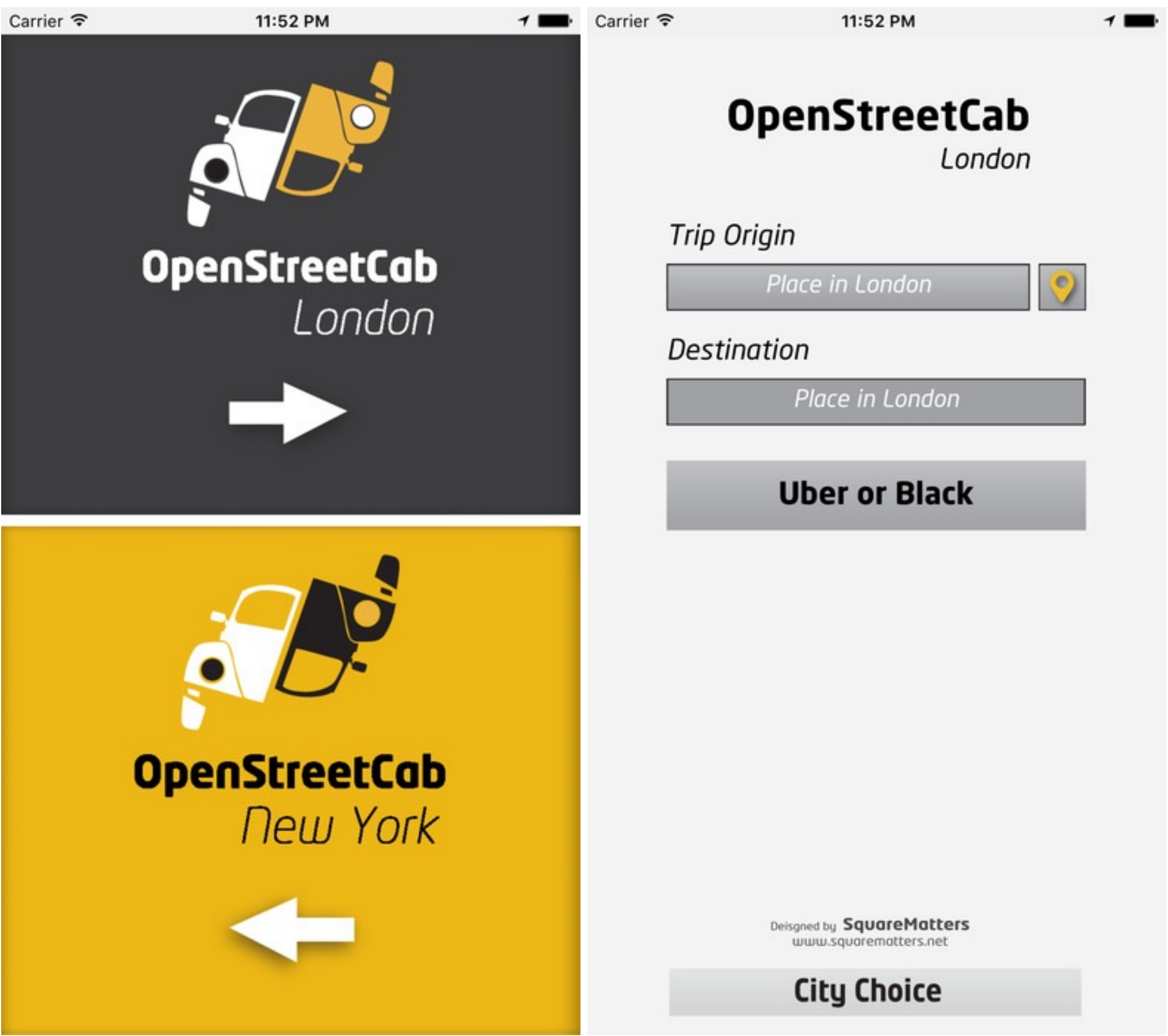}
    \end{subfigure}
    ~ 
    \begin{subfigure}[b]{.5\textwidth}
  \centering
    \includegraphics[width=0.8\columnwidth]{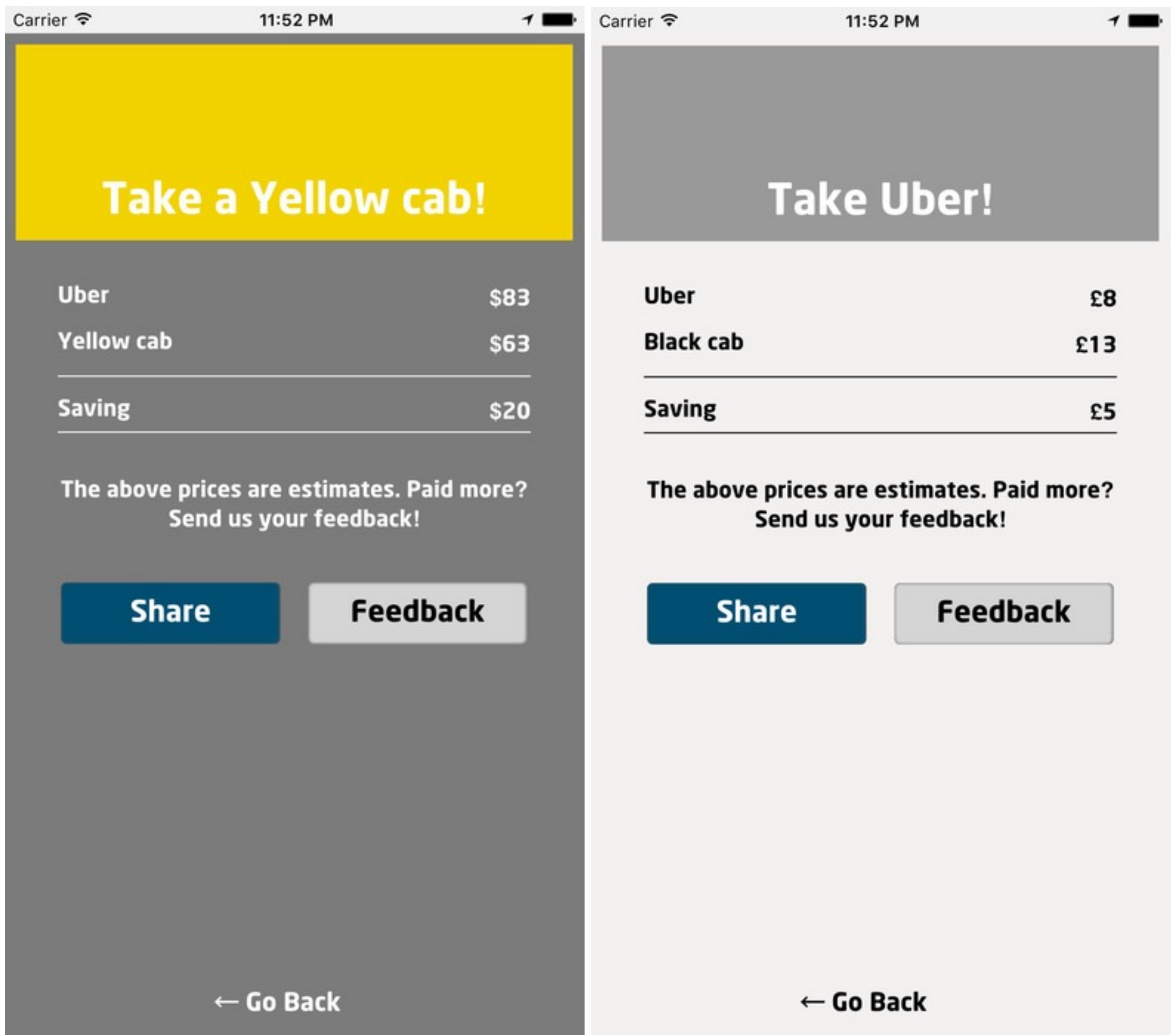}
    \end{subfigure}
  \caption{The application's user interface. From left to right, (a) city selection screen, (b) journey query submission screen and price estimation screens for (c) New York and (d) London.}~\label{fig:oscshot}
\end{figure}

\section{Application and System Design}
\label{sec:application}
We now present in detail OpenStreetCab's application logic and system architecture. After the initial screen where a user is prompted to choose a city (Figure\ref{fig:oscshot}.a), she is presented with a simple screen requesting origin and destination geographic coordinates of the imminent journey she is intending to take (Figure\ref{fig:oscshot}.b).
The app then returns price estimates on two major taxi providers and a recommendation 
on the cheapest one (Figure\ref{fig:oscshot}.c and \ref{fig:oscshot}.d). 

\subsection{User interface}
We have first launched the app in New York City in March 2015, and subsequently in London in the very beginning of January 2016. As mentioned above, users that install the app need first to select their city of interest (London or New York). Subsequently
in the journey query submission screen they can specify their trip's origin and 
destination. We provide two functionalities to enable user localisation: first, a button 
next to the origin input tab that automatically sets the origin address, given the user's geographic location  (through GPS / WiFi sensing), and, second, a text-input geocoding that parses user input 
and matches it to the most similar address name.

After setting the origin and destination addresses for a journey the user can press a button, `Uber or Yellow?' in New York or `Uber or Black' for London for comparison between Uber X and Yellow Cabs or Black Cabs respectively. This will push the input query to our server where Uber prices are compared to the competing local provider (see Section~\ref{sec:system} for the specifics on calculations). 
Next, the user is presented with a screen where price estimates are provided, including an indication on the price difference (`Savings'), with an additional projection of a colored header at the top of the screen clearly indicating the taxi provider for which the estimate is lower (e.g., Yellow for yellow taxis in New York). 
\begin{figure}
    \centering
    \includegraphics[width=\columnwidth]{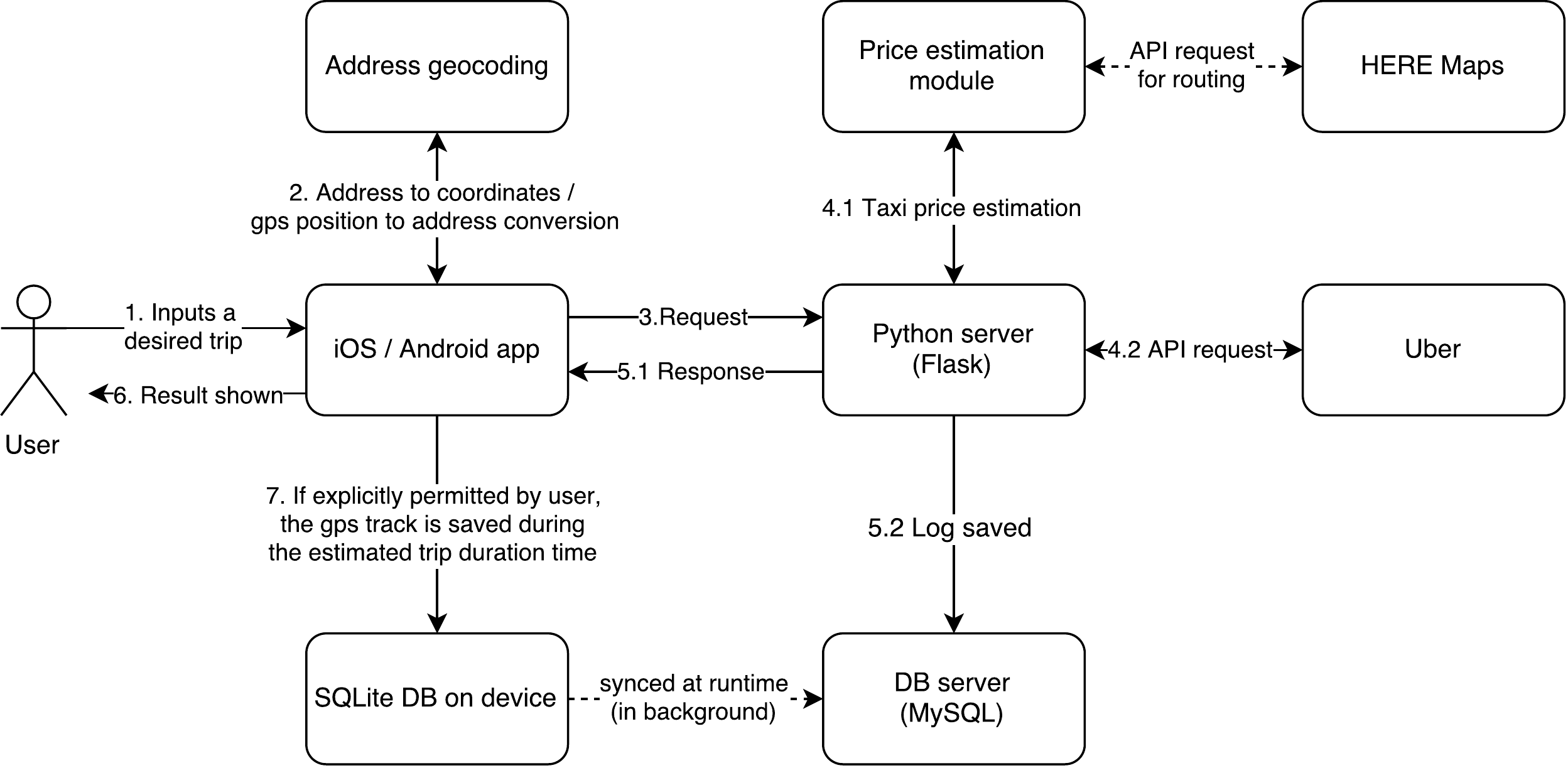}
    \caption{Information flow in the application.}~\label{fig:Diag}
\end{figure}
\subsection{How the App works}
\label{sec:system}
In addition to the user input, 
data is collected on the time of the user query:
a GPS sample of the user's current location and the application installation unique identification number.
The latter has been useful to associate users with submitted queries over time, as  
we required no registration information for our users. 
Once the user input  is gathered this is pushed to our servers for computation. 
We use a flexible architecture framework moving as much as possible of 
the application's logic to the server side. This approach avoids delays due to approval times required primarily by the App Store 
in case of minor modifications of the app or urgent bug corrections required (the App Store 
can sometimes take more than two weeks to approve a new version of a submitted app). 

Uber prices for the journey are collected through the Uber developer API~\cite{uberapi}.
The API returns two values, min and max, that define a price range for the costs of its Uber X service. Next, the mean estimate is calculated from these values, rounding to the closest integer value.
We chose to provide the mean as opposed to ranges, as in a list of a few providers it would be easier to compare on a single value as opposed to a range. 
Traditional taxi providers do not typically provide APIs on pricing. Instead, different taxi companies use different tariff schemes. We therefore combine information on tariffs for Yellow and Blacks Cabs in New York and London respectively, with routing information offered by HERE Maps~\footnote{\url{https://here.com/}}. HERE Maps return a shortest, in terms of time duration, routing path that is sensitive to traffic information the company gathers from a variety of sources. We then simulate the taxi's meter along the route and estimate the price of a journey according to the tariff information in each city. Black Cabs in London feature a more complex tariff logic~\footnote{\url{https://tfl.gov.uk/modes/taxis-and-minicabs/taxi-fares}} than the Yellow Taxi company in New York~\footnote{\url{http://www.nyc.gov/html/tlc/html/passenger/taxicab_rate.shtml}}. In principle, tariff schemes apply a flat cost
known as \textit{flag} in the beginning of the journey when passenger boards and then the price meter increases as a function of time and distance. For example, a rule may suggest that fare increases by a fixed ammount (e.g. $X$ U.S. Dollars) after $Y$ meters or $Z$ seconds (whatever comes first). HERE Maps returns the routes as a set of segments, technically referred in the system as \textit{manoeuvres}. For each route segment there is information on the length in kilometers and the typical driving time taken to drive on the segment. We exploit this information to increment the fare of the journey according to the tariff rules. Tariff rules depend also on time (e.g. morning versus night) and dates (e.g. holidays versus regular days) and we have integrated this aspect of pricing into OpenStreetCab as well. What is more, special destination or origin points such as airports or train stations can imply additional costs as well as costs that are specific to the route of the journey taken such as tolls. As currently there is no system that provides such information on routes, we have relied on keeping our system to date through manual labour and very critically on user feedback. 

An overview of the system's architecture is provided in Figure~\ref{fig:Diag}.
The client side component is handling the following: the user input and interactions that were described in the previous paragraph, location geocoding (including reverse geocoding) according to functionality provided in the corresponding platform (iOS or Android) and the output of the html-formatted response coming from the server.

The user input is pushed to a Python-based backend server, operating with Flask~\footnote{\url{http://flask.pocoo.org/}} microframework, which communicates with two price estimation APIs to retrieve estimates based on the origin 
and destination geographic coordinates provided through the user input. 
User journey queries and pricing data are pushed and saved on the MySQL database server after being
collected temporarily in an SQLite database on the client side. 
The data is also stored on the MySQL database server. 

\section{User Growth and Application Statistics}
\label{sec:analysis}
In this section we provide an overview of our application usage statistics. 
We then introduce an analysis on queries submitted by its users focusing on
price comparisons and differences across the two cities in which we have launched. 
\begin{figure}[t!]
\centering
  \includegraphics[width=\columnwidth]{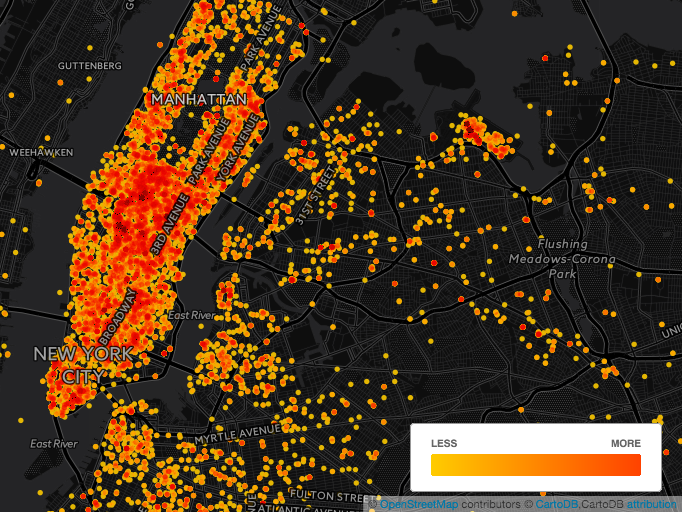}
  \caption{Query distribution in New York considering journey origin.}~\label{fig:newyorkmap}
\end{figure}

\subsection{Data Collection}
\begin{figure}[t!]
    \centering
    \includegraphics[width=0.8\columnwidth]{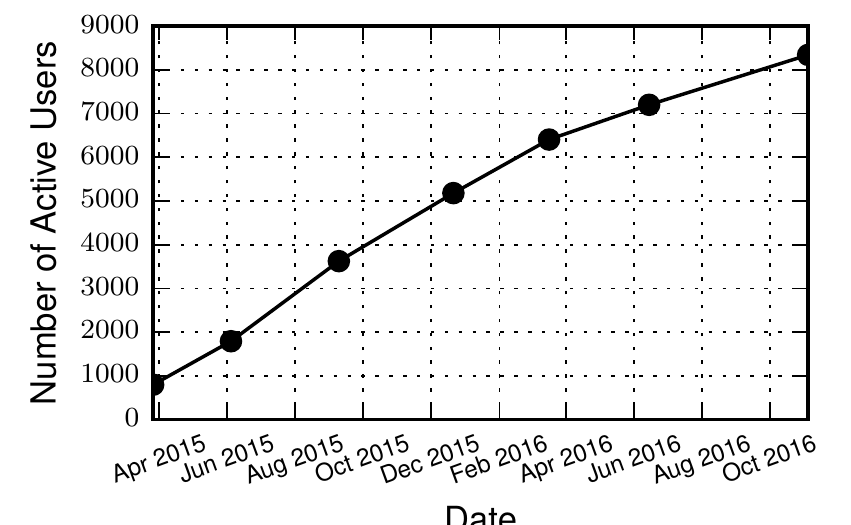}
    \caption{User growth over time considering users with at least one query since install.}~\label{fig:usergrowth}
\end{figure}

Overall, since the launch of the app in March 2015, more than $13, 000$ users have installed it in the two cities with around $75\%$ of all installs taking place on an iOS platform and the rest on Android. In Figure~\ref{fig:usergrowth} we present the number of OpenStreetCabs users that have submitted at least one journey query. Approximately $8000$ users have submitted a query, more than $70\%$ of those ever installed the application. Usage trends vary seasonally, but the number of total users with at least one query every three months is in the range of $1500$ to $2000$. The average number of queries per user is equal to $3.12$ with almost $350$ user having submitted $10$ queries or more.
In Table~\ref{tab:appstats} we provide a summary of the statistics by city together with the total
number of journey queries submitted. Regarding the number of queries, in New York there 
were a total of $25,804$ queries submitted to our server. The geographic dispersion of user queries 
is shown on the map of New York in Figure~\ref{fig:newyorkmap}, where the heatmap shows the spatial
variations in query frequency. As expected most activity is concentrated in Manhattan with occasional
hotspots in peripheral areas that include New York's La Guardia airport.  

We have measured an average saving of $8$ U.S. Dollars per journey
considering the mean difference between provider prices in each query. This corresponds to total potential savings of almost $206,000$ U.S. Dollars for the app's users assuming that they always choose the cheapest provider.
The number of queries in London are $3,371$ with potential savings of $12,405$ British Pounds on an average price difference of $3.68$ GBP (Great British Pounds). While this number may
not be reflective of the real amount of money saved, since users may not pick always the cheapest
provider (e.g. due to personal criteria regarding service quality), its scale is indicative of the potential financial impact that similar apps can have on the taxi market. 

\begin{table}[t!]
  \centering
  \begin{tabular}{l r r r}
    & & \multicolumn{2}{c}{\small{\textbf{Application Statistics}}} \\
    \cmidrule(r){3-4}
    {\small\textit{City}}
    & {\small \textit{iOS installs}}
      & {\small \textit{Android installs}}
    & {\small \textit{Queries}} \\
    \midrule
    New York & 9340 & 3095 & 25804 \\
    London & 1030 & 345 & 3371\\
  \end{tabular}
  \caption{Summary of application statistics across platforms and cities.}~\label{tab:appstats}
\end{table}
\subsection{Data analysis}
We now focus on taxi journey price estimates across providers. 
Tariffs differ between providers not only due to different vehicle maintenance and insurance costs. Black Cabs in London are historically luxurious, offering even wheel chair accessibility. 
Another difference between the two cities' traditional providers is that Yellow Cabs in New York are operated by large companies that own large fleets of those, while in London Black Cab drivers typically are the owners of the vehicle as well. One cannot drive a Black Cab 
without extensive professional training (more details on driver training are provided in Section~\nameref{sec:experiment}).
Overall these differences can imply different operational costs and point to the direction of potential variations in journey prices as well. 
Licensing, if any, is typically less complex and much less costly for Uber drivers. 

Black Cabs in London operate on a
tariff scheme~\footnote{\url{https://tfl.gov.uk/modes/taxis-and-minicabs/taxi-fares/tariffs}} that determines pricing depending
on both time and distance, following the rule \textit{For each $X$ meters or $Y$ seconds (whichever is reached first) there is a charge of $Z$ GBP} with the actual numbers depending on the time of the day and current meter price. The minimum charge is $2.4$ GBP (almost $3.55$ U.S. Dollars). For New York the initial charge is $2.5$ U.S. Dollars with extra $50$ cents charged every 5th of a mile or the same amount for
$50$ seconds in traffic or when the vehicle is stopped. Uber X applies a minimum fare of $5$ GBP in London\footnote{\url{https://www.uber.com/cities/london/}} and a charge of $0.15$ GBP per minute and $1.25$ GBP per mile. For New York the base fare is $2.55$ dollars, 
with a charge of $35$ cents per minute and an additional charge of $1.75$ USD per mile. Due to surge pricing however, 
Uber X fares can increase with the total amount being multiplied by a \textit{surge multiplier}. As previous works
have shown, surge pricing can happen rather frequently and can be highly sensitive in spatio-temporal terms with changes happening
across distances of a few meters or a few seconds~\cite{chen2015peeking}. Note also that Uber has been changing their tariffs rather frequently as opposed to traditional providers that typically change their tariffs rather slowly. For instance in New York City yellow cab fares increased in 2012 after eight years~\cite{yellowfares}.
\begin{figure}[t!]
    \centering
    \includegraphics[width=0.8\columnwidth]{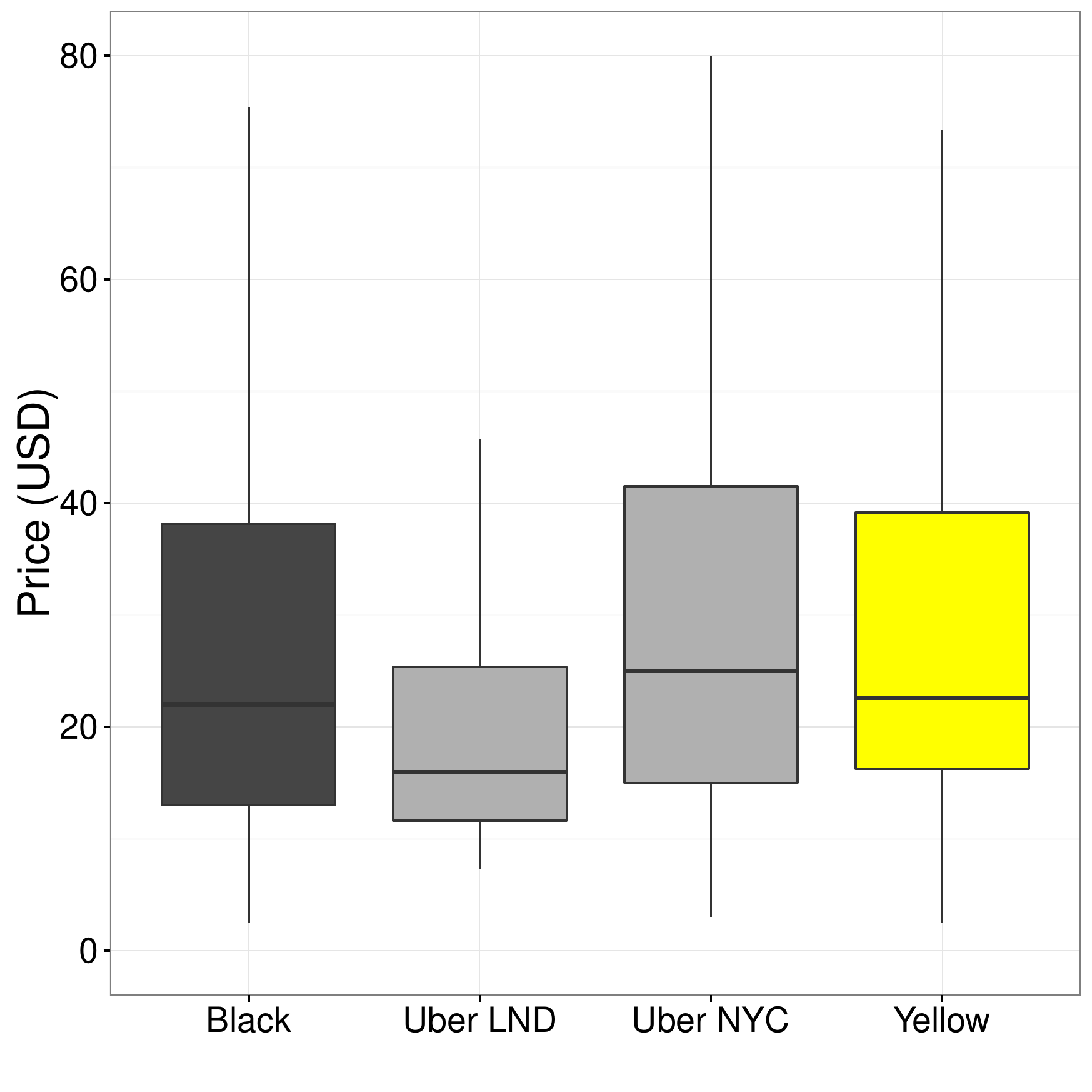}
    \caption{Box-and-whisker plot showing price query distributions for Black Cab in London, Uber in London, Uber in New York and Yellow Cab in New York (from left to right).}~\label{fig:cityprices}
\end{figure}
\begin{figure}[t!]
    \centering
    \includegraphics[width=0.8\columnwidth]{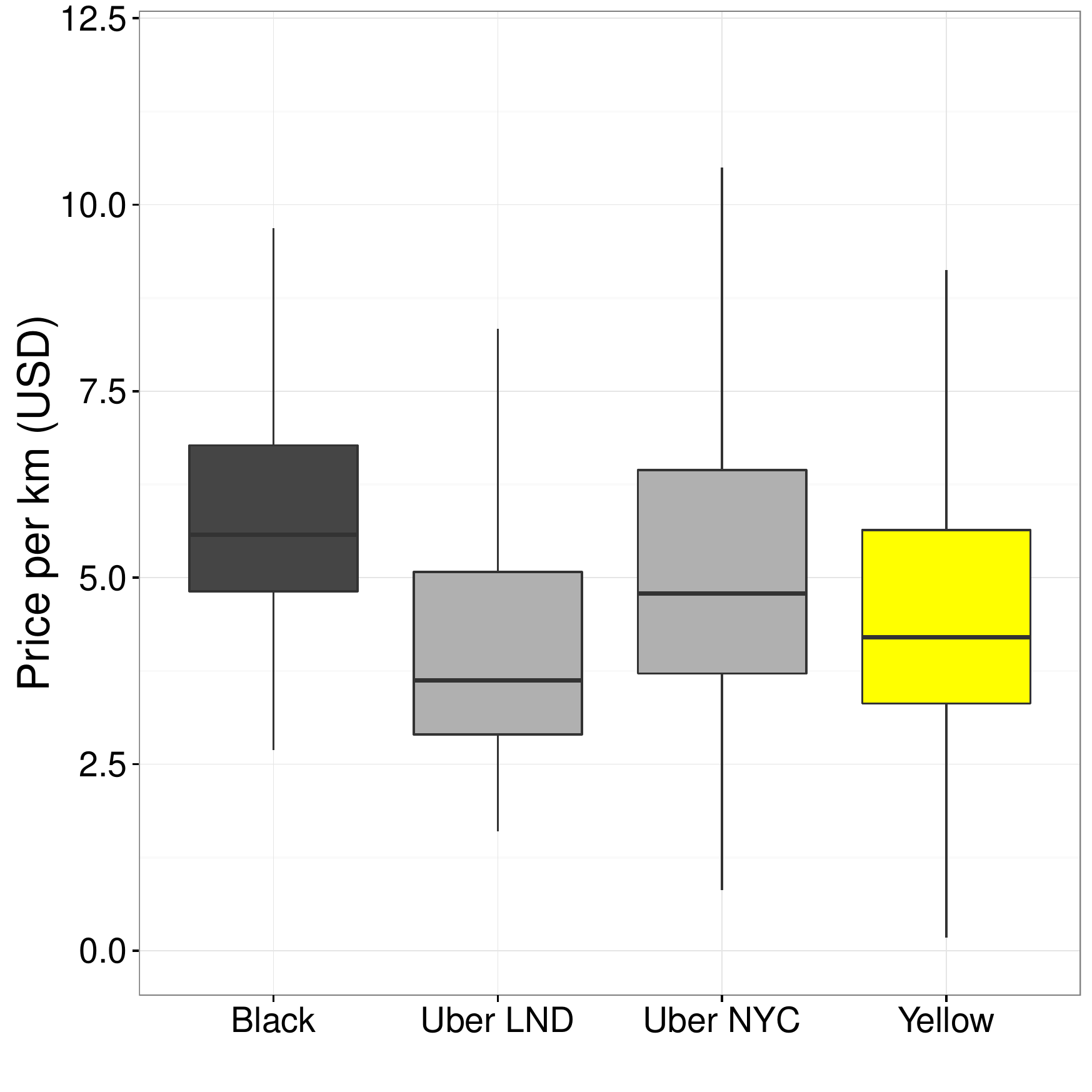}
    \caption{Box-and-whisker showing price distributions normalised by distance for Black Cab in London, Uber in London, Uber in New York and Yellow Cab in New York (from left to right).}~\label{fig:cityprices_normed}
\end{figure}

\begin{figure}[t!]
    \centering
    \includegraphics[width=0.8\columnwidth]{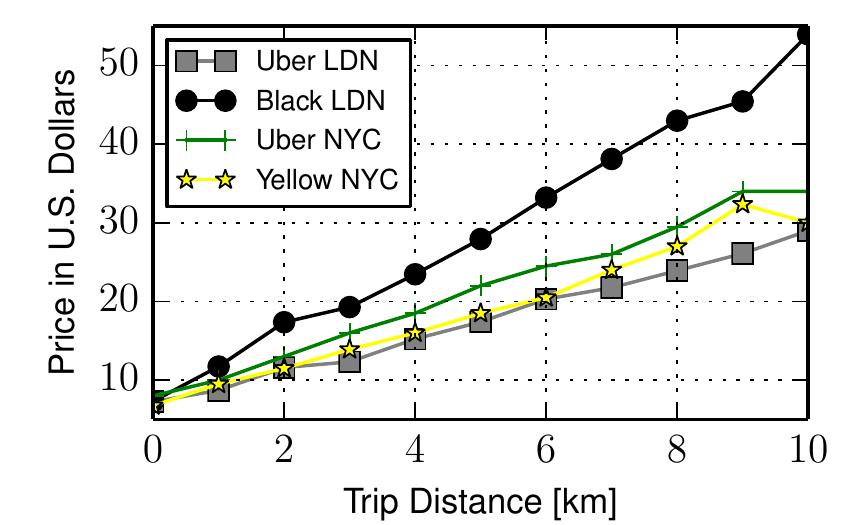}
    \caption{Prices versus journey distance for taxi providers in London and New York.}~\label{fig:pricesVsDistance}
\end{figure}

\begin{figure}[t!]
    \centering
    \includegraphics[width=0.8\columnwidth]{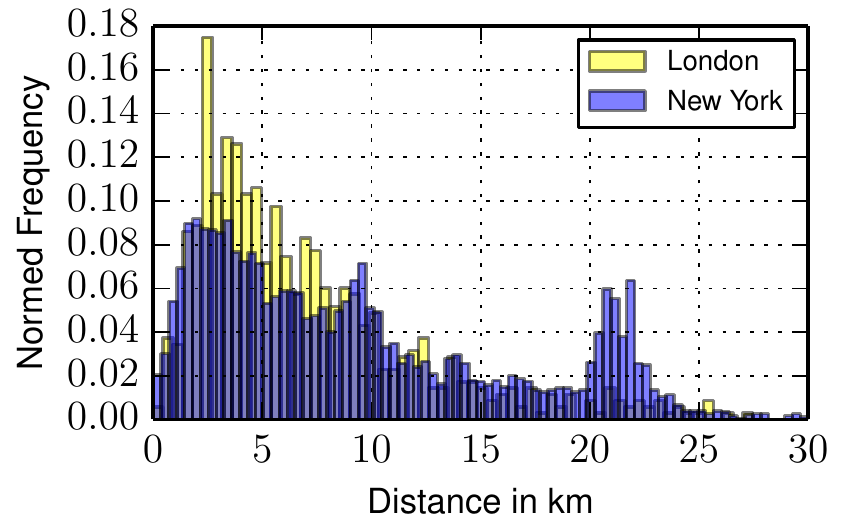}
    \caption{Geographic distance histograms of journey queries.}~\label{fig:querydists}
\end{figure}

The box-and-whisker plot, shown in Figure~\ref{fig:cityprices}, describes the distribution of price queries from our app split into quartiles. Each box represents the mid-quartile range with the black line in the middle representing the median of the distribution, while the ``whiskers" represent the top and bottom quartiles of the distribution. 
The median values are $25$ USD for Uber X in New York, $22.5$ USD for Yellow Cabs, $16.6$ USD for Uber X in London and $23.8$ USD for Black Cab in London, respectively. 
We note that while Uber X is on average more expensive in New York City as compared to the local provider, this is not the case for London where the service appears considerably cheaper. 
Even in the latter case however, a surge multiplier of $1.5$ or more could translate to a more expensive trip. As the tariffs discussed above would suggest, Black Cab should be more expensive compared to the New York providers. 
This is confirmed in Figure~\ref{fig:cityprices_normed} where prices are normed by distance 
showing that Black Cabs are more expensive on a per km basis. As also implied through normalisation by distance users in New York  tend to make longer journey queries. 
In Figure~\ref{fig:pricesVsDistance} the mean journey price is shown in relation to distance as measured through the app's user queries. We note 
a steeper increase over distance for Black Cabs. While they are clearly getting more expensive as distance grows it is worth noting that in practice long journeys beyond $5$ km are relatively rare. In terms of the geographic distances for journey queries submitted through our application, the two cities appear to show similar trends as shown in Figure~\ref{fig:querydists} with a peak at small distances of $2$ or $3$ kilometers. However, in New York journey distances of approximately $20$ km are particularly common due to a large number of  queries submitted for journeys to and from JFK airport. 
Interestingly, Uber X is much cheaper in London than in New York. 
It is hard to explain this difference as it could relate to aspects of its pricing model or marketing reasons. For instance, lower prices could be due to higher availability of drivers in relation to user demand. On the other hand, the company may have put forward a strategy of lower fares in London sacrificing perhaps short term revenue in order to increase demand and its share in a more competitive market. 

\section{Taxi Experiments in the Wild}
\label{sec:experiment}
So far we have observed high variability between the prices of taxi providers in the two cities through our app.
In order to validate the app's price estimates however, we need
to collect ground truth evidence on taxi journeys. We therefore ran a three day experiment on the ground in the city of London. 
Beyond validating prices, we took this opportunity to measure journey times and routing behavior for the two competing taxi providers in the city of London: Uber -- focusing on their basic Uber X service, and the city's traditional Black Cab service. 
There are some well known differences between the two services which we take into consideration in analyzing the output of our experiments.
To acquire a license in London, Black Cab drivers need to attend a school that takes about three years to complete and pass \textit{The Knowledge}~\cite{TheKnowledge} test that thoroughly examines the ability of drivers to know by heart the whereabouts of a large number of streets and points of interest in central London. Notably, medical tests on these drivers have suggested that their training and profession results to a larger number of cells in the hippocampus region of the brain which is the region that hosts the spatial navigation mechanism for mammals~\cite{maguire2000navigation}. Another advantage of the Black Cab service is that they are licensed with Transportation for London, which means they can use bus lanes across the city. On the other hand, Uber drivers do not receive any special training and rely exclusively on their navigation system. These differences are noticeable to users of the two services in the city but no quantifiable data-driven insights exist on these differences until now. 
\begin{figure}[t!]
    \centering
  \centering
    \includegraphics[width=0.8\columnwidth]{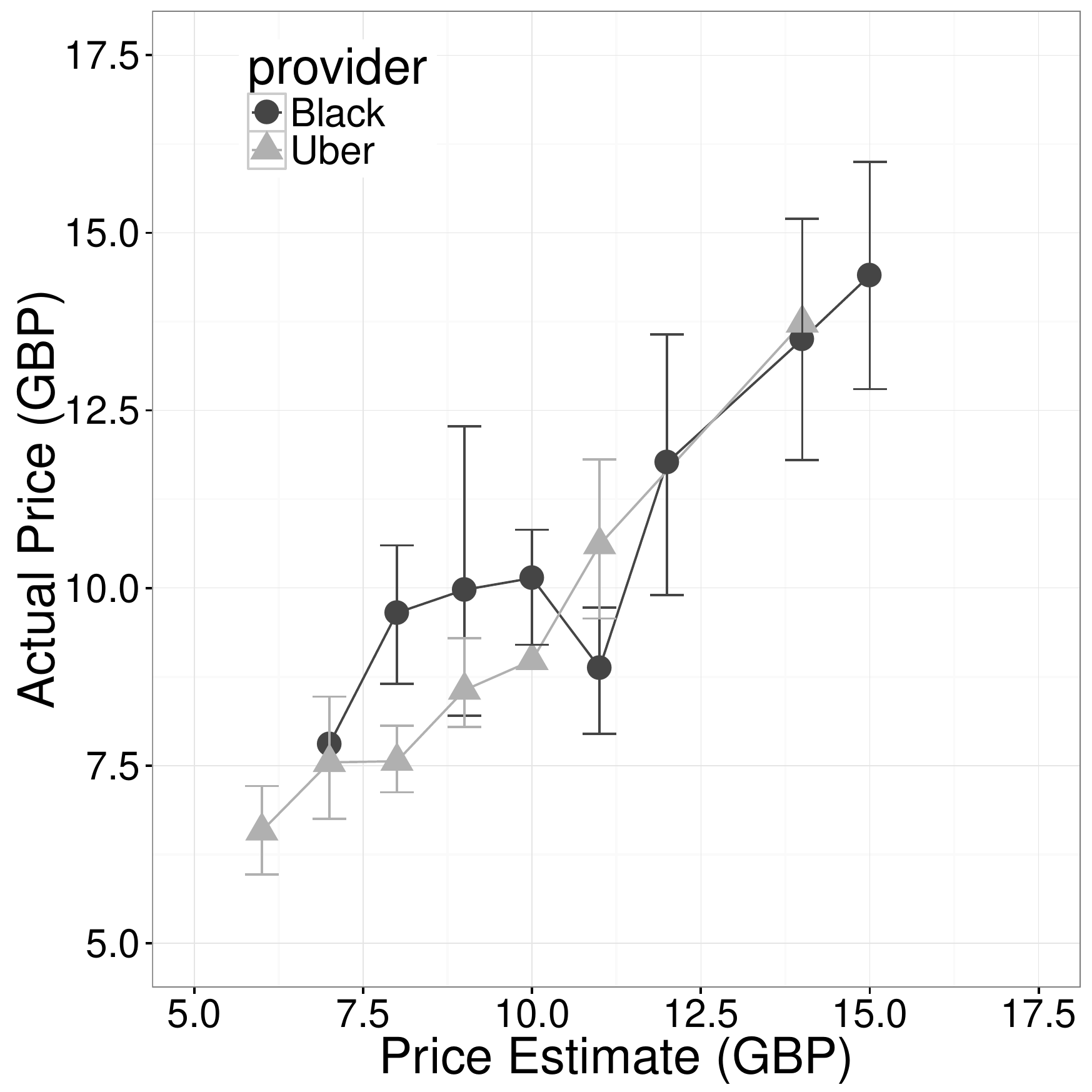}
    \caption{Application price estimates versus average actual amounts paid in London. The errors bars 
    correspond to standard errors.}~\label{fig:priceEstimates}
\end{figure}
\subsection{Experimental Setup and Conduct}
The experiment took place in London over three consecutive days in February 2016.
Two researchers performed $29$ side-by-side journeys comparing the prices, times and routes between Uber X and Black Cab in London. 
Using an in-built route tracking functionality (not yet enabled for standard users), the GPS coordinates of trajectories followed by each provider 
were recorded along with their respective timestamps, start and stop journey times and price estimates from the app. 
Black Cab and Uber receipts were collected in the end of each trip so estimates could be compared to actual prices. 

The journeys selected for the experiment were based on popular user queries for the app in London. Each researcher commenced the journey from the same geographic origin to the same destination at approximately the same time taking either Uber or Black Cab. 
Temporal synchronicity is very important in this setting, especially in central areas of the city, where traffic conditions could change dramatically in a matter of a few of minutes. Whereas absolute temporal synchronicity is almost impossible in a realistic context, to minimise temporal differences in start times, an Uber X was booked through the Uber app at a location where it was easy to pick up a Black Cab. That was either possible at locations where Black Cab ranks were present or busy intersections where it was possible to hail one easily. At peripheral areas where it was not easy to hail a Black Cab, the application Hailo was used~\cite{hailo}. The latter allows the booking of Black Cabs in London offering a very similar functionality to Uber. 
Special attention was also paid to geographic coverage with the intention of covering central busy parts of the city, but also peripheral areas in the North, South, West, and East of central London.  Overall, over the course of three days, operating roughly between 11am and 10pm, in total more than $300$ km were covered.

\subsection{Incorporating Driver Feedback}
While map APIs provide information on shortest routes given origin and destination information, also taking into account real time traffic information as HERE Maps or Google Maps do, these systems are configured to reflect the behavior of a regular car drivers. 
We hypothesize that this may not necessarily reflect the routing and driving capabilities of professional drivers.
Furthermore, many Black Cab drivers mentioned that they do not use a navigation system as they know whereabouts in the city well through training and experience. In light of this possibility, following drivers' feedback, we have introduced a reduction coefficient to the price estimates of Black Cabs, assuming they are able to typically route faster in a territory they are trained and experienced in driving. 

Through the application's feeback mechanism, drivers got in touch to report issues and reflect on the app. In some cases, we received metered validation tests run by a driver for a number of routes.
\begin{framed}
Your taxi fares estimate always seems high. There are a few regular journeys I inputted and did the same with quite a few jobs and every time I have come in under your taxi estimate.

3 examples were:
New Kings Rd, SW6 to Grovesnor Crescent, SW1, usually \pounds 9, your app says 11 \pounds,
Belgrave Sq, SW1 to Heathrow Terminal 5 usually \pounds 55, your app says 60 \pounds,
New Kings Rd to Canary Wharf usually \pounds 38 when your app says 40 \pounds. (Ross, Driver, London).
\end{framed}
Following his suggestion in addition to others 
we decided to reduce estimates for Black Cabs by adding a multiplier coefficient of $0.9$ (reduction by $10$\%). As we validate in Section~\ref{sec:experiment}, the correction has improved predictions overall. 

We then asked the same driver to run a few more tests and provide feedback.
\begin{framed}
I compared your estimates with about 10 jobs I done today and your pretty much spot on. A couple were pounds 1-2 over, a couple were under and some were on the button!

Another thing I thought of is to maybe let people know that these estimates are for the Taxi day rate (rate 1) and for Uber without a surge price in effect.

For taxis there are 3 rates, rate 1 from 06:00-20:00, rate 2 from 20:00-22:00 and rate 3 from 22:00-06:00. To be honest there isn't much difference between rates 1 and 2, but rate 3 does make it a bit more expensive.
(Ross, Driver, London).
\end{framed}

Next, we show results of fare estimate prediction with and without using the reduction coefficient. 
\subsection{Experimental Results}
\paragraph*{Price Estimates}
\begin{figure*}[t!]
    \centering
    \begin{subfigure}[b]{.44\columnwidth}
    \includegraphics[width=\columnwidth]{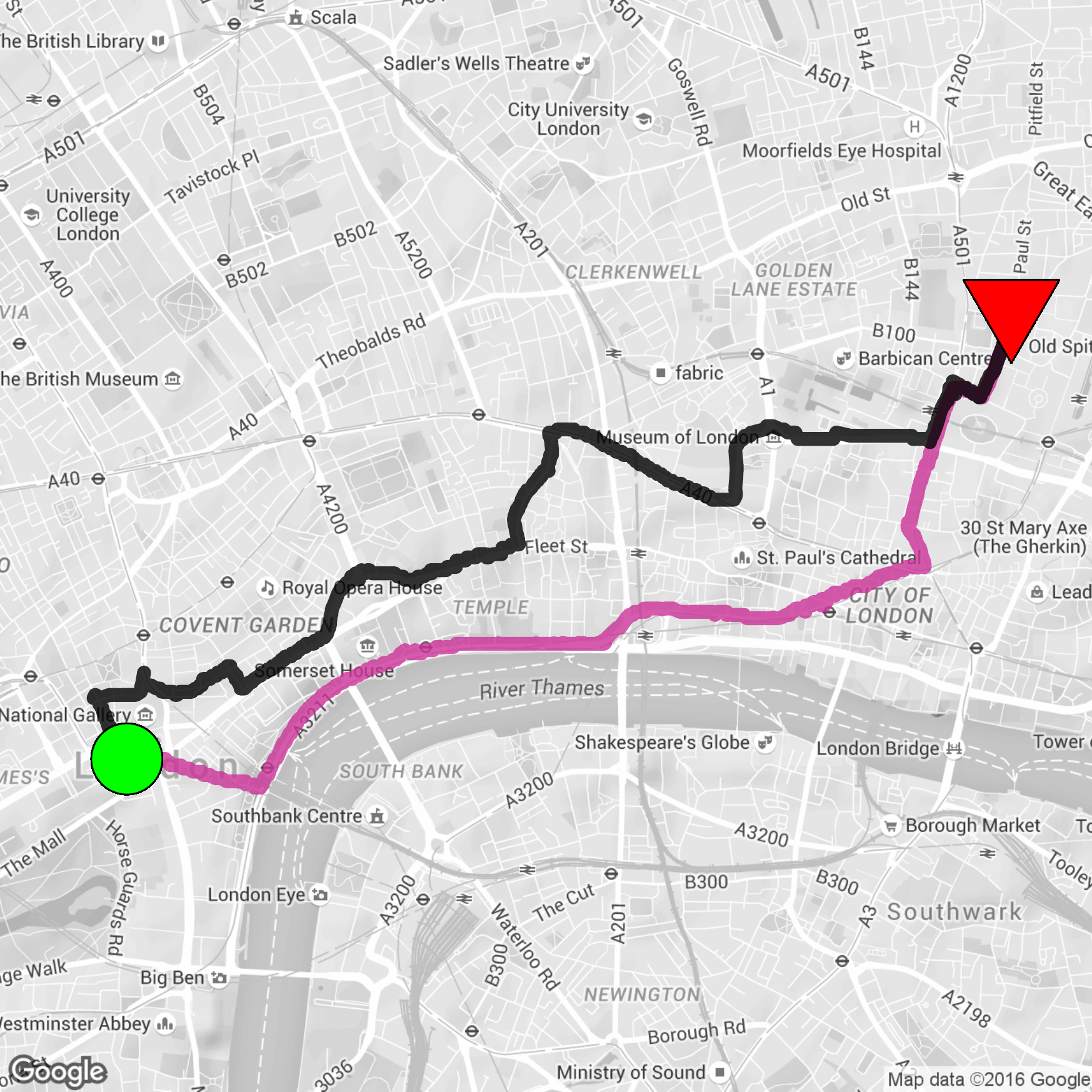}
    \caption{Trafalgar Square - Old Street}~\label{fig:trafalgar}
    \end{subfigure}
    ~ 
    \begin{subfigure}[b]{.44\columnwidth}
  \centering
    \includegraphics[width=\columnwidth]{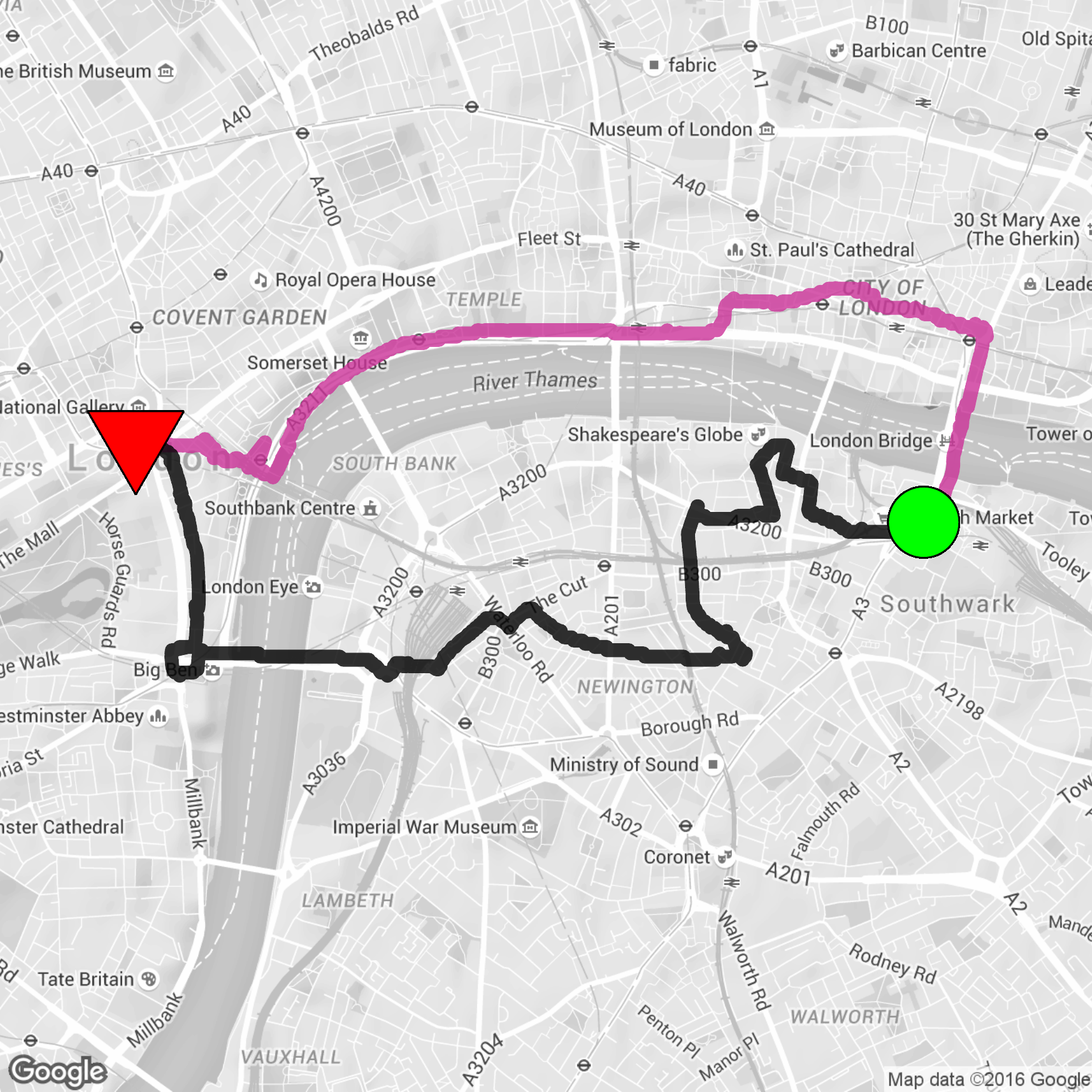}
    \caption{London Bridge - Trafalgar Square}~\label{fig:waterloo}
    \end{subfigure}
      \centering
    \begin{subfigure}[b]{.44\columnwidth}
    \centering
    \includegraphics[width=\columnwidth]{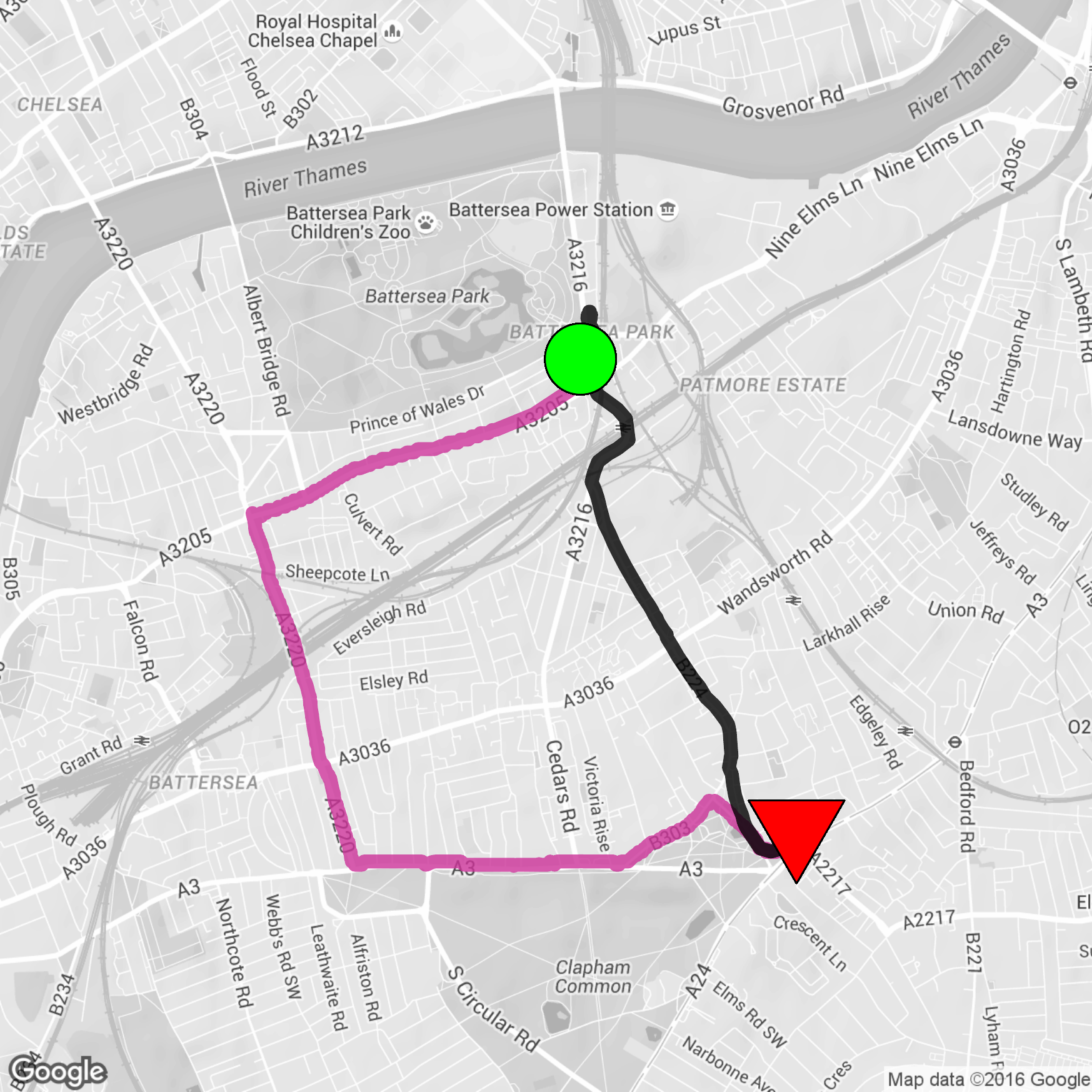}
    \caption{Battersea Park - Clapham Common}~\label{fig:clapham}
    \end{subfigure}
    ~ 
    \begin{subfigure}[b]{.44\columnwidth}
  \centering
    \includegraphics[width=\columnwidth]{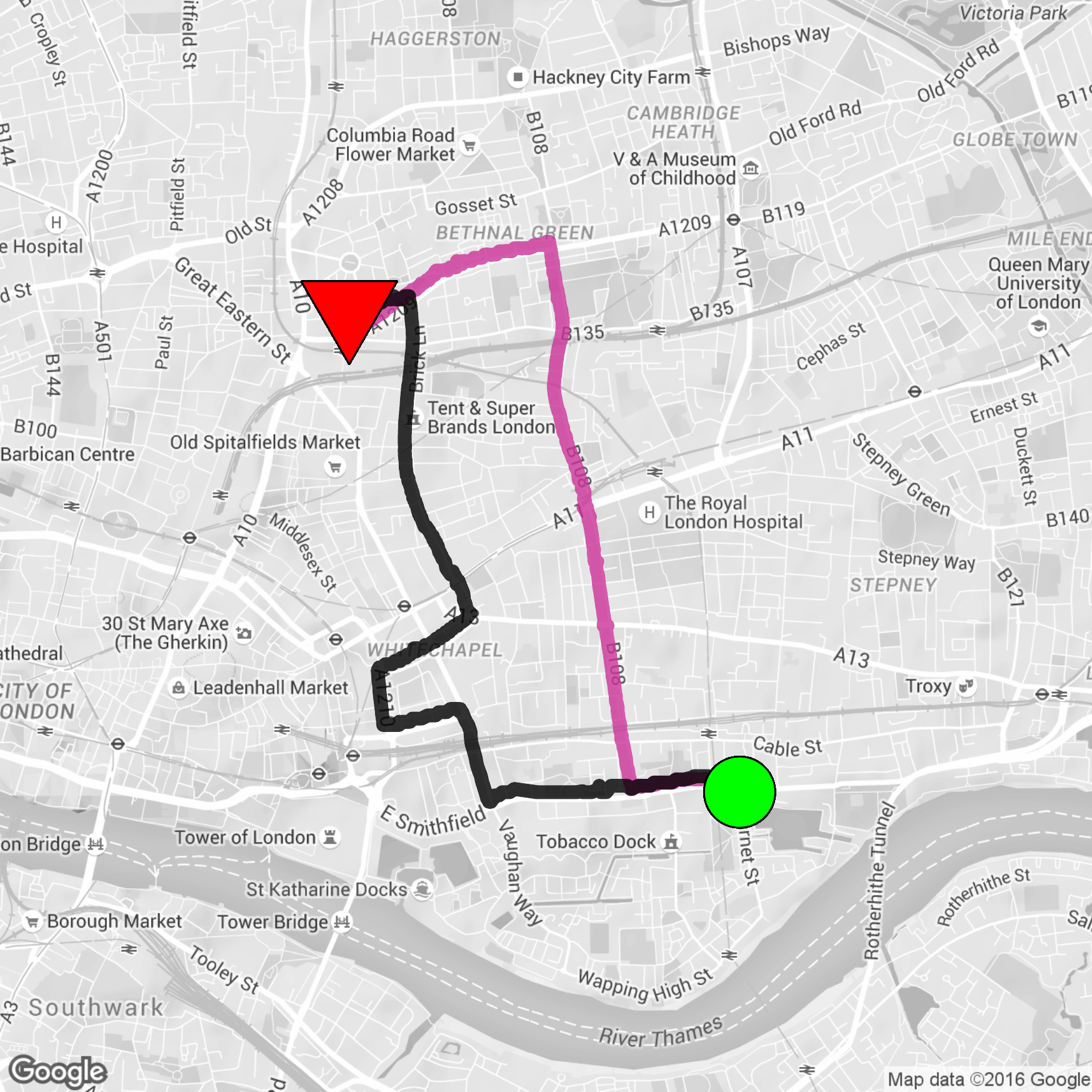}
    \caption{Shadwell Basin - Shoreditch}~\label{fig:shoreditch}
    \end{subfigure}
    \caption{Taxi provider trajectories in four areas of London. Black Cab in black color and Uber X in pink. Origins are marked with a Green circle and Destinations with a Red triangle. }\label{fig:journeymaps}
\end{figure*}
\begin{table*}[t!]
  \begin{tabular}{l r r r r r r r}
    \multicolumn{7}{c}{\centering\small{\textbf{Journey Estimation Statistics}}} \\
    {\small\textit{Provider}}
    & {\small\textit{Max Abs Diff}}
    & {\small \textit{Mean Abs Diff}}
    & {\small \textit{Std Diff}}
    & {\small \textit{Max \% Dev}} 
    & {\small \textit{Mean \% Dev}} 
    & {\small \textit{Std \% Dev}} 
    & {\small \textit{Pearson's $\rho$}} \\
    \midrule
    Black Cab & $4.4$ & $0.06$ & $1.96$ & $0.45$ & $0.15$ & $0.11$ & $0.81$\\
    Black No feedback & $3.5$ & $-1.04$ & $2.01$ & $0.59$ & $0.18$ & $0.16$ & $0.81$\\
    Uber X & $3.31$ & $0.03$ & $1.08$ & $0.32$ & $0.10$ & $0.06$ & $0.83$\\
  \end{tabular}
  \caption{Accuracy of price estimates for Black Cabs, Black Cabs prior to receiving feedback from drivers
  and Uber X. Mean, standard deviations and maximum price difference are shown in terms absolute (abs) or 
  fractional (\%) values.}~\label{tab:expstats}
\end{table*}

For every journey with an Uber X or a Black Cab in the experiment, we have compared our application's estimate measured
as described in Section~\nameref{sec:analysis} against the actual price charged by the provider. In Figure~\ref{fig:priceEstimates} we plot the mean actual price charged for a given price
estimate and the corresponding standard error. We consider the overall estimates for both providers
satisfactory, yet deviations exist. For Black Cabs deviations were higher for journeys that cost between
$7$ and $9$ GBP. In the case of Uber, estimates tend to be more stable, however, deviations
still remain.
We provide more detail, in Table~\ref{tab:expstats}, where we show statistics in terms of absolute and percentage values on the maximum price difference in GBP (column \textit{Max Abs Diff}), mean price difference (column \textit{Mean Abs Diff}) and the standard deviations (column \textit{Mean Std Dev}) between actual and price estimates for three estimate scenarios: Black Cab after incorporating driver feedback on top of the original price estimates, reducing price estimates by $10$\%, the case when feedback is ignored (no feedback) and the estimates provided by the Uber API on Uber X. In the case of Uber X 
price estimates have deviated from actual ones on average by $10$\%. For Black Cabs 
estimates deviated on average by $15\%$. It is worth noting that prior to introducing a reduction coefficient of $0.9$ in response to driver feedback estimates were deviating more, on average by $18\%$. The Pearson's $\rho$ correlations between estimated and actual price values are $0.81$ for Black Cabs and $0.83$ for Uber. While the introduction of a reduction coefficient may appear overly simplistic at first glance, deploying more complex strategies can in fact yield worse estimates. The heterogeneity of routing decisions is very high in complex urban street networks which typically unfold in large cities and in this setting we found that simple engineering decisions are more robust than introducing a complex logic in the price estimation engine.  

In general, the variations in price estimates may be due to inherent differences between predicted and actual routes, urban congestion changes that are not accurately picked up by navigation systems or the ability of drivers to route themselves differently to what is predicted by navigation systems. 
Driver input, as we have observed in the previous section, may reflect aspects of Black Cab driver behavior that are not picked by modern routing APIs. In fact, none of the Black drivers used a navigation system during the experiment and they are likely to pick different routes than what a computer system would suggest. However, we can see that even in the case of Uber where drivers typically follow the company's navigation system predictions cannot be perfect. 
We note that Uber provides ranges of price estimates (min and max values) for prices. We chose to use only a single average value (mean) to make direct comparisons between providers easier. Users may expect and tolerate some variation between predicted and actual prices and the average here serves as an indicator on how much cheaper a provider may be compared to another. In future versions of the app, aside from the possible inclusion of price ranges, that could be inspected for instance with a click on the provider price (in case a user is willing to access more details regarding pricing) other information could be added such as journey time estimation. We discuss a related analysis in the next paragraph.

\paragraph*{Provider Comparison}
\begin{figure}[t!]
\centering
  \includegraphics[width=0.8\columnwidth]{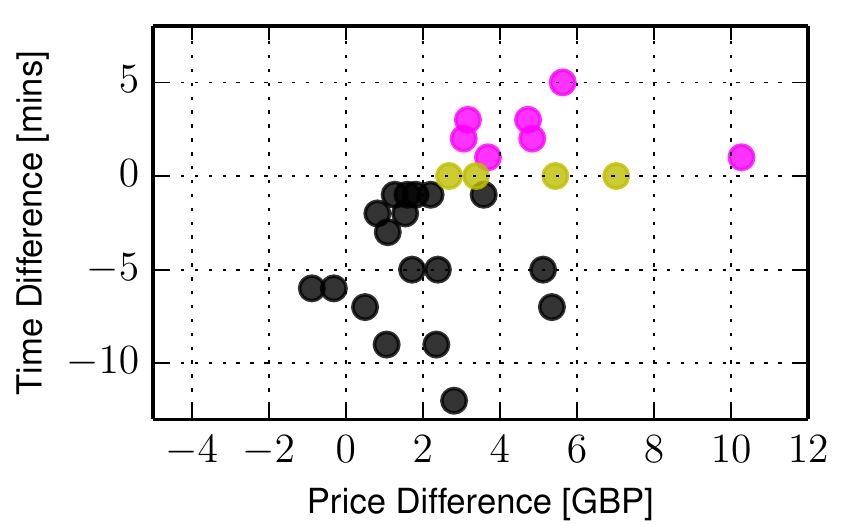}
  \caption{Price versus time differences where price difference is defined as Black Cab price minus Uber price in GBP and time difference as Black Cab journey time minus Uber X journey time in minutes. Black colored circles correspond to faster journey times for Black Cabs, pink for Uber X and yellow for ties.}~\label{fig:PriceVsTime}
\end{figure}

We have empirically observed significant variations in terms of how the two providers compare in terms of actual and estimated prices, with routing choices being the most probable reason for these deviations.  In Figure~\ref{fig:journeymaps} we show four characteristic journeys where routes had very little geographic or no overlap at all between the two providers. Black cab drivers tend to take more complex routes in terms of picking side streets as opposed to larger main streets that are recommended more often by GPS navigation systems as part of shortest path routing.
As already implied in previous sections by tariffs applied Black Cabs were in general more expensive. Uber X would cost on average $74\%$ of a Black Cab's journey price. 

Nevertheless, Black Cabs were faster and took on average $88\%$ of an Uber's trip duration, where average journey time has been $14.06$ minutes Black Cabs and $16.34$ minutes for Uber (Uber or Black Cab waiting times excluded). Out of the $29$ journeys, Black Cabs were faster in $18$ cases, there were $4$ ties and Uber X was faster in $7$ instances. Figure~\ref{fig:PriceVsTime} presents a scatter plot
reflecting the relationship between price and time differences. The faster Black Cabs have been, as one  would expect by definition of the pricing schemes that depend on time in addition to route length, the smaller the price difference. Further, when Black Cabs have been faster, in almost half of the occasions ($10$ times) they have been faster by $5$ minutes or more. 
\begin{figure}[t!]
    \centering
    \begin{subfigure}[b]{.5\textwidth}
  \centering
    \includegraphics[width=0.8\columnwidth]{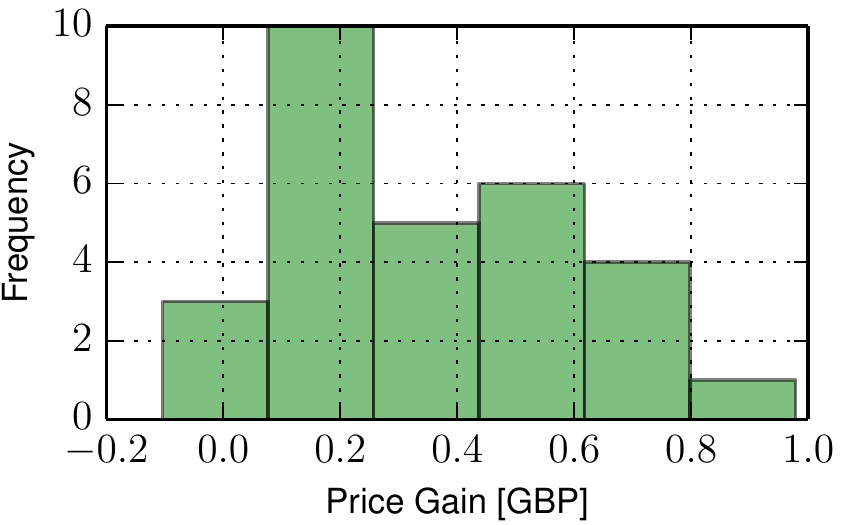}
    \caption{Frequency distribution for price gains for all journeys. Price gains are higher for Uber.}~\label{fig:priceGains}
    \end{subfigure}
    ~ 
    \begin{subfigure}[b]{.5\textwidth}
  \centering
    \includegraphics[width=0.8\columnwidth]{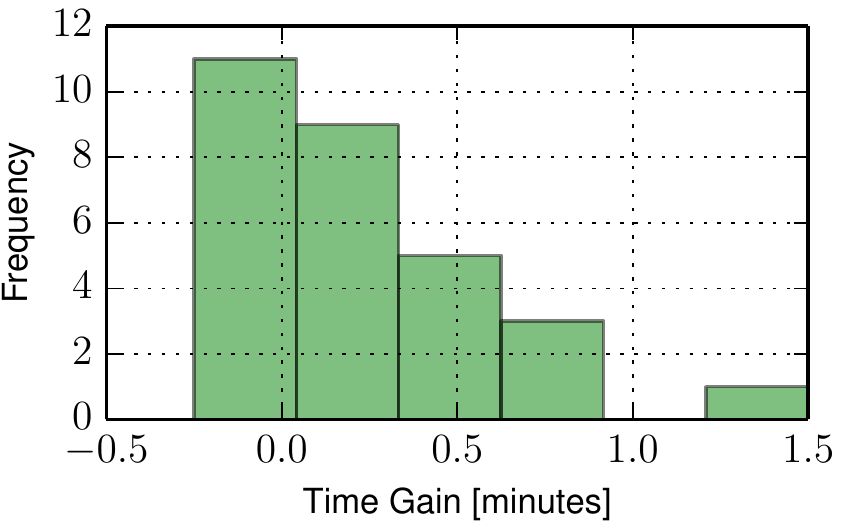}
    \caption{Frequency distribution of time gains for all journeys. Time gains are higher for Black Cabs. }~\label{fig:timeGains}
    \end{subfigure}
    \caption{Price and time gains for $29$ side-by-side journeys between Black Cab and Uber X in London.}\label{fig:pricetimegains}
\end{figure}
To better understand the price and time differences across journeys and providers we define the relative gains of the two variables as: 
\begin{equation}
PriceGain = \frac{PriceUber-PriceBlack}{PriceBlack}
\end{equation}
for prices, and for times as:
\begin{equation}
TimeGain = \frac{TimeUber-TimeBlack}{TimeBlack}
\end{equation}
The corresponding frequency distributions are shown in Figures~\ref{fig:priceGains}~and~\ref{fig:timeGains} respectively. 
These contrasting results between price and time gains
point to a clear trade-off between time and price when considering the choice of a provider. From a user perspective, should they be in a hurry to catch the next train or a meeting, according to these results, Black Cab would appear to be 
a safer bet. Should they just be willing to save money on the particular journey then Uber X could be favored.


\paragraph*{The Impact of Urban Density}
Throughout the experiment we noticed that Black Cabs tend to be more time effective in the urban core of the city. This advantage would become less clear in journeys taking place towards more peripheral areas of the city. To empirically explore this intuition we characterized routes in terms of their average place density. We exploit Foursquare's venue database to do so. Foursquare is a local search service which provides a semantic location API~\footnote{\url{https://developer.foursquare.com/}}, allowing us to retrieve the number of businesses in an area and therefore is a proxy to the urban density of a particular area. For every coordinate sampled for a route of a provider, we defined a $200$ meter radius around it and counted the number of Foursquare places in vicinity, considering almost a set of $40,000$ venues in London becoming available through the services venue API. 
We then took the mean across all GPS points and across the two providers. 
Formally, we characterise the average urban density of a trip as:
\begin{equation*}
Trip\_Density = \frac{1}{|T|} \frac{\sum_{i=1}^{|T|} P(x=lng_i, y=lat_i, r=200m)}{\pi r^{2}}
\end{equation*}
where $T$ is the union set of the two provider trajectories made of GPS coordinates encoded as latitude and longitude values, and $P$ is a function that returns the number of Foursquare places that fall within a disc area, given as parameters a
radius $r$ set equal to $200$ meters, a geographic center represented by latitude $lat_i$ and longitude $lng_i$ coordinates, when considering a given point $i$ in set $T$. The unit of measurement is number of places
per square kilometer (hence the division by the area size $\pi r^{2}$). 

\begin{figure}[t!]
    \centering
  \centering
    \includegraphics[width=0.8\columnwidth]{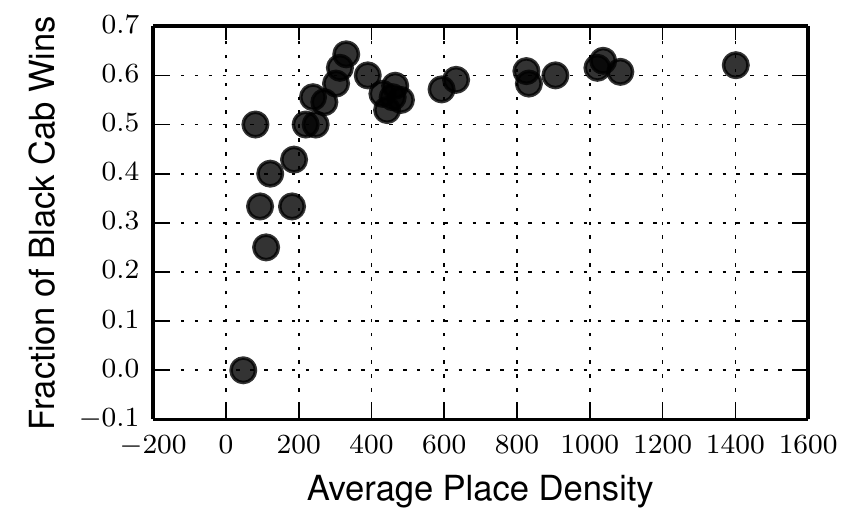}
    \caption{Fraction of Black Cab wins in terms of faster journey times for different place density values. 
    To extract the fraction value we count the number of Black cab wins over the total number of journeys that 
    feature a mean place density smaller or equal to a given value. The unit of density measurement is number of places
per square kilometer. }~\label{fig:routeDensityFraction}
\end{figure}

In Figure~\ref{fig:routeDensityFraction} we plot the fraction of Black Cab wins, counted as faster journey
achieved, versus the total number of trips that have a density smaller or equal to a given value x. We can observe, albeit the noise due to a small number of samples, that as density
values increase so does the relative cumulative probability Black Cab being faster. 
This result provides an indication that Black Cab drivers are especially effective in parts of the city where urban complexity in terms of urban congestion, street network and population density rises. In this setting navigation systems may be less effective in terms of reflecting actual traffic in real time and being able to provide quicker routes for drivers. 

\paragraph*{Discussion, limitations and future directions}
The empirical findings presented in this section provide novel insights not only on differences in the service characteristics between taxi providers, but also on differences in routing behavior between drivers relying on navigation systems to reach a destination, versus drivers who have been trained for years in wayfinding in the city. We show how experienced human navigators are able to choose alternative routes that can improve journey times, especially in the city center, where urban complexity increases. The results highlight the importance 
of integrating journey time as a significant economic factor in taxi and urban transport recommendations, but also point to potential weaknesses of navigation systems when those are used in dense and congested
urban environments.
 In future versions of our app we plan to integrate journey time comparisons together with that of prices and help users make more informed choices on the provider that fits their journey preferences
best.

From the point of view of experimental conduct there is considerable space for improvement. The 29 rides with each provider correspond to a limited sample and is only a first modest step towards understanding driver routing behaviour and provider service quality. Rerunning the experiment more times would not be sustainable from a financial and time cost perspective. We have therefore enabled crowdsourcing as a solution to scaling data collection on routes of different providers through tracking user trajectories. While the latter approach lacks the viewpoint of direct and controlled comparison between different taxi providers, it has the advantage of enabling the collection of a larger number of route samples for a given origin-destination pair. This could shed light, for instance, on  heterogeneities that may exist in terms of driving and routing behaviour over a well defined network of streets.

\section{Conclusions}
In this paper we have described our experience with the development and deployment of a price comparison mobile app for taxi rides.
The app was deployed in the wild in two cities and we show how the feedback received from both users and drivers drove further app updates and validation tests. The main lesson learned from the deployment and the feedback has been the importance of driver experience in route finding: our study has given ample evidence of this. The inclusion of these factors into a route finding system or even simply in an app like ours is not trivial and the object of our future work. Moreover, 
in future work we intend to introduce further crowdsourcing in terms of route selection and user experience (e.g., journey times, driver behavior).

\small
\bibliographystyle{plain}
\bibliography{biblio}

\end{document}